\documentclass[reprint, amsmath,amssymb,aps,prl,nofootinbib]{revtex4-2}

\usepackage{enumitem}
\setlist[enumerate]{itemsep=0mm}
\usepackage{bm}
\usepackage{color}
\usepackage{graphicx}
\usepackage{hyperref}
\usepackage{braket}
\usepackage{bbm}
\usepackage{tikz}
\usepackage{nicefrac}

\newcommand{\be}{\begin{equation}}
\newcommand{\ee}{\end{equation}}
\newcommand{\ba}{\begin{aligned}}
\newcommand{\ea}{\end{aligned}}

\newcommand{\bs}[1]{\boldsymbol{\mathbf{#1}}}

\begin{document}
\title{Universality in the tripartite information after global quenches}
\author{Vanja Mari\'c}
\affiliation{%
 Universit\'e Paris-Saclay, CNRS, LPTMS, 91405, Orsay, France%
}
\author{Maurizio Fagotti}
\affiliation{%
 Universit\'e Paris-Saclay, CNRS, LPTMS, 91405, Orsay, France%
}

\begin{abstract}
We consider macroscopically large 3-partitions $(A,B,C)$ of  connected subsystems $A\cup B \cup C$ in infinite quantum spin chains 
and study the R\'enyi-$\alpha$ tripartite  information $I_3^{(\alpha)}(A,B,C)$.  
At equilibrium in clean 1D systems with local Hamiltonians it generally vanishes. A notable exception is the ground state of conformal critical systems, in which $I_3^{(\alpha)}(A,B,C)$ is known to be a universal function of the cross ratio $x=|A||C|/[(|A|+|B|)(|C|+|B|)]$, where $|A|$ denotes $A$'s length.
We identify different classes of states 
that, under time evolution with translationally invariant Hamiltonians, locally relax to states with a nonzero (R\'enyi) tripartite  information, which furthermore exhibits a universal dependency on $x$.
We report a numerical study of $I_3^{(\alpha)}$ in systems that are dual to free fermions, propose a field-theory description, and work out their asymptotic behaviour for $\alpha=2$ in general and for generic $\alpha$ in a subclass of systems. This allows us to infer the value of $I_3^{(\alpha)}$ in the scaling limit $x\rightarrow 1^-$, which we call ``residual tripartite information''. 
If nonzero, our analysis points to a universal residual value $-\log 2$ independently of the R\'enyi index $\alpha$, and hence applies also to the genuine (von Neumann) tripartite information.  
\end{abstract}

\maketitle

The concept of entanglement was introduced to distinguish quantum systems from classical ones~\cite{Einstein1935Can,Bell1964On}. With the development of quantum information theory~\cite{Ingarden1976Quantum,Cerf1998Information}, such a peculiarity of the quantum world was recognised as a resource, and the quantification of the entanglement became a key question~\cite{Horodecki2009Quantum}. Various measures of entanglement have then been put forward.   
Besides their original purpose, the tools studied in quantum information attracted the attention of the community working on quantum many-body systems. It was indeed realised that the entanglement measures unveil universal properties.  
A famous example in 1D is the von Neumann entropy of an interval in the ground state of a (conformal) critical system, which has a typical logarithmic growth with the length, proportional to the central charge of the underlying conformal field theory (CFT)~\cite{Holzhey1994Geometric,Calabrese2004Entanglement,Korepin2004PRL}. Another quantity that attracted some attention (unfortunately under different names) is the \emph{tripartite information}, which contains more information about the underlying CFT\cite{Caraglio2008Entanglement,Furukawa2009Mutual,Calabrese2009Entanglement}; but not only that. 
In 2D  the tripartite information was shown to be sensitive to topological order and renamed for that reason ``topological entanglement entropy''~\cite{Kitaev2006Topological}. Concerning higher dimensions, we mention that Ref.~\cite{Casini2009Remarks} investigated the tripartite information in generic quantum field theories (QFT) (see also Ref.~\cite{Agon2022Tripartite}), among which those with holographic duals hold a special place~\cite{Hayden2013Holographic}.
More recently, a type of tripartite information was  proposed as a diagnostic of scrambling~\cite{Hosur2016Chaos,Schnaack2019Tripartite,Sunderhauf2019Quantum}.

The tripartite information is defined as~\cite{Cerf1998Information} 
\be
I_3(A,B,C)=I_2(A,B)+I_2(A,C)-I_2(A,B\cup C)
\ee
where $I_2(A,B)=S(A)+S(B)-S(A\cup B)$ denotes the mutual information and $S(A)\equiv S_1[\rho_A]=-\mathrm{tr}[\rho_A\log \rho_A]$ is the von Neumann entropy of subsystem $A$ with density matrix $\rho_A$. It is defined so as to cancel the extensive and the boundary contributions to the entropies. Moreover, just as the mutual information quantifies the extensiveness of the von Neumann entropy, so  $I_3(A,B,C)$ quantifies the (bi)extensiveness of the mutual information after fixing one of the subsystems. 
We focus on the case in which $A$, $B$, and $C$ are adjacent intervals in an infinite spin chain and assume that their lengths are asymptotically large:
$$
    \begin{tikzpicture}[scale=0.21]
    \draw[black,line width=0.6pt] (0,0) to (36,0);
    \foreach \x in {1,...,35}
    \filldraw[ball color=blue!20!white,opacity=0.75,shading=ball] (\x,0) circle (6pt);
    \node[anchor=south] at (10.5,0.5) {$A$};
     \node[anchor=south] at (20,0.5) {$B$};
      \node[anchor=south] at (28.5,0.5) {$C$};
    \draw[fill=blue,opacity=0.3, rounded corners = 2,thick] (3.65,-0.5) rectangle ++(12.75,1);
     \draw[fill=blue,opacity=0.3, rounded corners = 2,thick] (16.65,-0.5) rectangle ++(6.75,1);
 \draw[fill=blue,opacity=0.3, rounded corners = 2,thick] (23.65,-0.5) rectangle ++(8.8,1);
    \end{tikzpicture}
$$
In noncritical 1D systems at equilibrium (and with clustering properties), $I_3(A,B,C)$ approaches $ 0$ as the lengths approach infinity (we are not aware of exceptions). A more interesting behavior is observed in the ground state of a critical system with a low-energy CFT description,
where conformal symmetry forces $I_3(A,B,C)$ to be a function of the cross ratio $x=\frac{|A||C|}{(|A|+|B|)(|B|+|C|)}$~\cite{Calabrese2009Entanglement}.
That is to say, the limit
\be\label{eq:scal_lim}
I_3(A,B,C)\xrightarrow[{|A||C|/[(|A|+|B|)(|B|+|C|)]=x}]{|A|,|B|,|C|\rightarrow\infty}G(x)
\ee
exists, is universal, and can in principle be computed within the underlying CFT. 
As a matter of fact, some difficulties of the calculation have not yet been overcome. 
The most important intermediate results concern the R\'enyi generalisation of the tripartite information, which we will refer to as R\'enyi-$\alpha$ tripartite  information and indicate by $I_3^{(\alpha)}(A,B,C)$. This has the same definition of $I_3(A,B,C)$ with the von Neumann entropy replaced by the R\'enyi-$\alpha$ entropy $S_\alpha[\rho]=\tfrac{1}{1-\alpha}\log\mathrm{tr}[\rho^\alpha]$.
Ideally, the tripartite information is recovered in the limit $\alpha\rightarrow 1^+$.
Provided that the intervals $A,B,C$ are adjacent\footnote{If the intervals are disjoint, the CFT prediction depends instead on three cross ratios --- see, e.g., Refs~\cite{Coser2014On,DeNobili2015Entanglement}}, in a CFT also the R\'enyi-$\alpha$ tripartite information depends only on the cross ratio $x$, i.e., $I_3^{(\alpha)}(A,B,C)= G_\alpha(x)$, and, in some theories, it has been computed exactly for generic integer $\alpha>1$~\cite{Furukawa2009Mutual,Calabrese2009Entanglement1,Calabrese2011Entanglement,Ares2021Crossing}. Just as $G(x)$ does for $I_3$, so $G_\alpha(x)$ describes the scaling limit of $I_3^{(\alpha)}$ in any spin chain with the same underlying CFT~\cite{Fagotti2010disjoint,Alba2010Entanglement,Alba2011Entanglement,Fagotti2012New,Coser2016Spin}. 

In this Letter we investigate the limit of \emph{infinite time} of $I_3^{(\alpha)}$ after global quenches with \emph{local} Hamiltonians.
The entropies are known to become extensive~\cite{Alba2017Entanglement,Bertini2022Growth,Zhou2020Entanglement}, and often the system exhibits typical features of thermal states~\cite{srednicki94,Polkovnikov2011Colloquium,Gogolin2016Equilibration}.  
We discuss when $I_3^{(\alpha)}$ should be expected not to vanish (in contrast to thermal states) and argue that it captures universal properties. 
We point out, in particular, that a residual tripartite information with a discrete value $-\log 2$ can emerge. 

\paragraph{The model.}

We focus on two classes of noninteracting spin chains. The first is the generalised XY model~\cite{Suzuki1971The}, which is mapped into free fermions by a  Jordan-Wigner transformation
$
\bs a_{2\ell-1}=\prod_{j<\ell}\bs\sigma_j^z\ \bs\sigma^x_\ell$, $
\bs a_{2\ell}=\prod_{j<\ell}\bs\sigma_j^z\ \bs\sigma^y_\ell
$,
where $\bs a_\ell$ are Majorana fermions satisfying $\{\bs a_\ell,\bs a_n\}=2\delta_{\ell n}\bs I$ and $\bs \sigma^\alpha_\ell$ are Pauli operators. 
The most studied models of this class are described by the quantum XY Hamiltonian~\cite{Lieb1961} in a transverse field
\be\label{eq:XY}
\bs H=\sum\nolimits_\ell J_x\bs\sigma_\ell^x\bs\sigma_{\ell+1}^x+J_y\bs\sigma_\ell^y\bs\sigma_{\ell+1}^y+h\bs\sigma_\ell^z\, .
\ee 
This includes, in particular, the XX model ($J_x=J_y$) and the transverse-field Ising model ($J_y=0$).

The second class of systems is mapped into free fermions by the Kramers-Wannier transformation 
$\bs\tau_\ell^x=\prod_{j\leq \ell}\bs\sigma_j^x$, $\bs\tau_\ell^y=(\prod_{j< \ell}\bs\sigma_j^x)\bs\sigma_\ell^y\bs\sigma_{\ell+1}^z$ 
(for the sake of clarity, we have used a different notation, $\bs\tau_\ell^\alpha$, for the Pauli operators),
followed by the aforementioned Jordan-Wigner transformation. An example is the dual XY model~\cite{Zadnik2021The,Fagotti2022Global}
\be\label{eq:dualXY}
\bs H=\sum\nolimits_\ell\bs\tau_{\ell-1}^x(J_x\bs I-J_y\bs\tau_\ell^z)\bs\tau_{\ell+1}^x\, .
\ee
Hamiltonians like \eqref{eq:dualXY} posses  semilocal conserved operators~\cite{Fagotti2022Global}, which enable symmetry-protected topological order after global quenches~\cite{Fagotti2022Nonequilibrium}. 
Since the tripartite information was recognised as an indicator of topological order in 2D~\cite{Kitaev2006Topological}, how $I_3$ behaves after a quench in this second class of models is a compelling question.  

\paragraph{Homogeneous quench from a critical ground state.}

\begin{figure}[!t]
\includegraphics[width=0.45\textwidth]{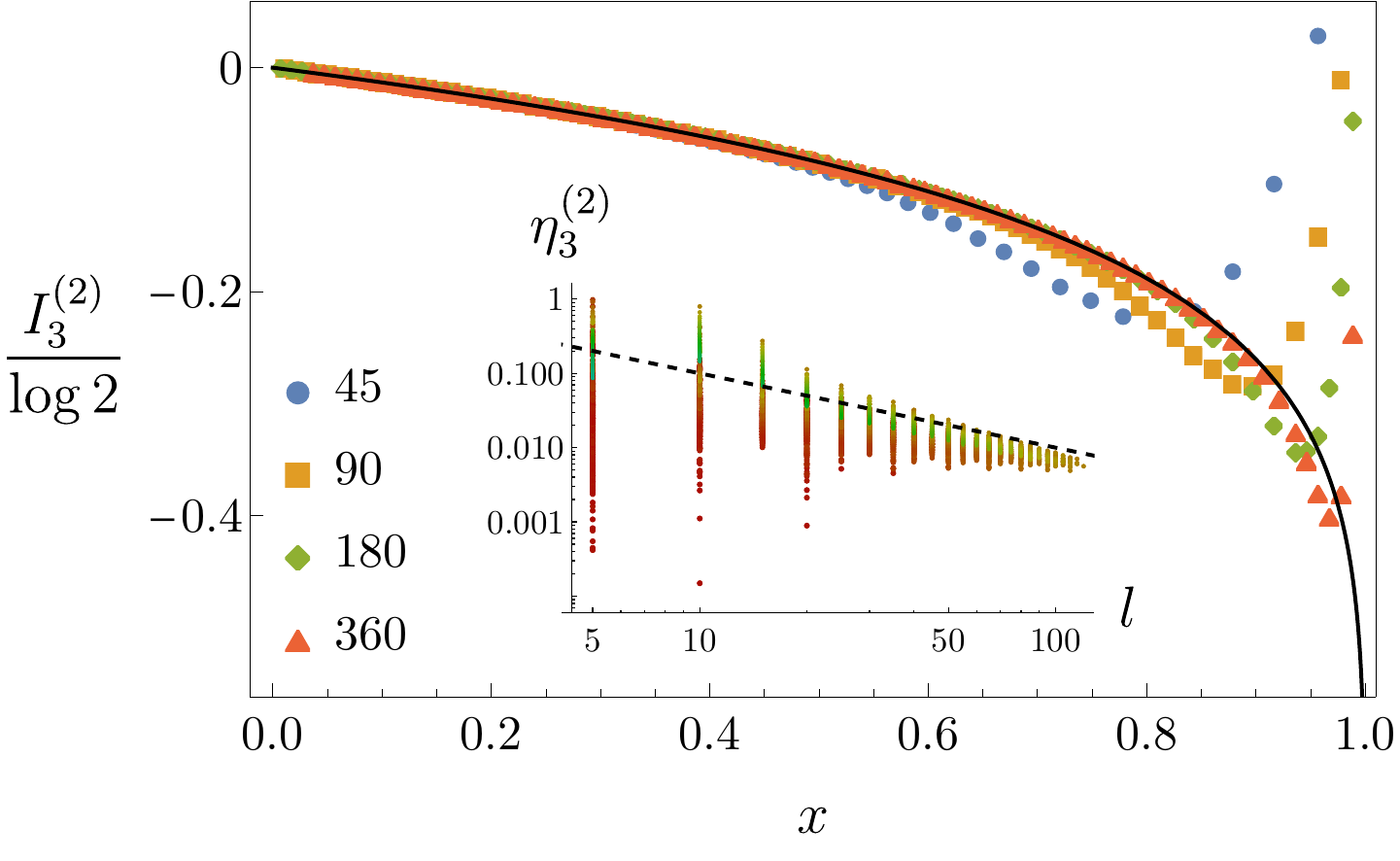}
\caption{R\'enyi-2 tripartite information for the quench in the XY model $(J_x,J_y,h): (1,1,1)\rightarrow (1,0.5,0.5)$ for $|A|=|C|\in\{45,90,180,360\}$ and variable $|B|$. The solid curve is prediction~\eqref{eq:G2conj}. Inset: the relative error $\eta_3^{(2)}$ as a function of $l=\min(|A|,|B|,|C|)$ for all configurations with lengths multiple of $5$ in the range $[5,120]$; the color and the size of the points varies linearly in $x$; the dashed line is a guide for the eye $\sim 1/l$.  }
\label{f:1}
\end{figure}

The first example we consider is the paradigm of global quench: the ground state of a translationally invariant Hamiltonian is let to evolve under a different translationally invariant Hamiltonian~\cite{Essler_2016}. We focus on generalised XY models. In generic situations $I_3^{(\alpha)}$ vanish (both at the initial time and) at infinite time after the quench (cf. Refs~\cite{Fagotti2010disjoint,Coser2014Entanglement}).
We are about to uncover exceptions when the initial state is the ground state of a conformal critical system.

In noninteracting models the corresponding central charge, $c$, is a multiple of $\frac{1}{2}$. Our numerical analysis shows that generically $I_3^{(\alpha)}$ are nonzero also at late times, independently of whether the post-quench Hamiltonian is critical or not, just provided that $c\geq 1$ in the initial state. This condition seems to be related to how slow the slowest spatial connected correlations decay. Specifically, after quenches from critical ground states there are 2-point correlation functions that decay with the distance as a power-law. 
In all the cases investigated with the initial state in the Ising universality class ($c=\frac{1}{2}$)  we find that the spin connected correlations in the stationary state do not decay more slowly than $1/r^4$, where $r$ is the distance; this is not enough to generate nonzero tripartite information. With $c\geq 1$ in the initial state we find instead that the slowest spin correlations generally decay as $1/r^2$, and $I_3^{(\alpha)}$ become nonzero (negative). Fig.~\ref{f:1} shows $I_3^{(2)}$ at late times after a quench in the XY model. Remarkably, $I_3^{(2)}$ remains a function of the cross ratio. This is observed also for larger values of $\alpha$ and other choices of the Hamiltonian parameters; in addition, the  data seem to approach curves that depend on few system's details~\cite{Maric2023UniversalityLong}.

\paragraph{Bipartitioning protocol.}

\begin{figure}
  \includegraphics[width=0.45\textwidth]{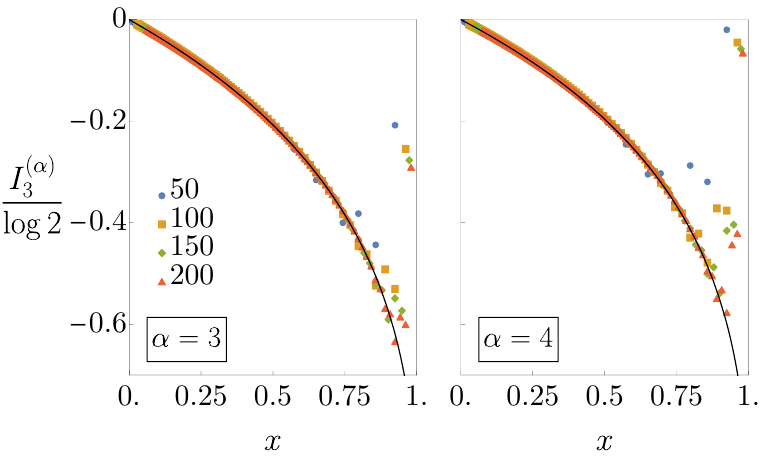}
  \caption{R\'enyi-$\alpha$ tripartite information at infinite time after the  quench from  $\ket{\ldots\uparrow\uparrow\downarrow\downarrow\ldots}$ under the XY Hamiltonian~\eqref{eq:XY} with $(J_x,J_y,h)=(1,0.5,0)$ for $|A|=|C|\in\{50,100,150,200\}$ and variable $|B|$. The solid curves are predictions~\eqref{eq:Galphapred}.}
 \label{f:2}
\end{figure}

Another type of global quench that has attracted a lot of attention is the time evolution after joining  two globally different states~\cite{Alba2021Generalized}. We consider here the basic case in which the initial state consists of a domain of spins aligned along $z$  joined with a domain of spins aligned in the opposite direction. If we take the Hamiltonian of the XY model with $J_x\neq J_y$ \emph{the quench is global}, indeed the initial state is locally different from any excited state of the Hamiltonian. We stress that the initial state is \emph{not} critical  
and we are not aware of any CFT description of the infinite time limit. 
Fig.~\ref{f:2} shows the R\'enyi-$3$ and R\'enyi-$4$ tripartite information in the nonequilibrium steady state emerging at infinite time. In agreement with the previous discussion, the slowest connected correlations decay as $1/r^2$ and we obtain nonzero $I_3^{(\alpha)}$. Again, the latter become functions of the cross ratio and seem to remain so even in more sophisticated bipartitioning protocols~\cite{Maric2023UniversalityLong}. In the XY model considered here the asymptotic curves do not even seem to depend on $J_x$ and $J_y$.

\paragraph{Quench with symmetry-protected topological order.}

We consider time evolution under \eqref{eq:dualXY} of two initial states: (a), the product state with all spins aligned in the $z$ direction (cf. Ref.~\cite{Fagotti2022Nonequilibrium}), (b), the same as (a) with one spin flipped (cf. Ref.~\cite{Fagotti2022Global}). In case (a), $I_3^{(\alpha)}$ vanish. In case (b), $I_3^{(\alpha)}$ become nonzero and seem to approach the same curves as in the domain-wall quench above. Since the latter is dual to case (b)~\cite{Fagotti2022Global}, our analysis suggests that $I^{(\alpha)}_3$ are not affected by the Kramers-Wannier duality. 
A more detailed analysis of global quenches in this kind of systems is reported in Ref.~\cite{Maric2023universality2}.

\paragraph{Towards universality.}

All systems investigated with nonzero $I_3^{(\alpha)}$ exhibit extensive entropies with logarithmic corrections, which can in turn be traced back to the presence of discontinuities in the ``filling function'' (or Fermi weight) $\vartheta(k)$ \cite{Alba2009Entanglement,Ares2014}. We remind the reader that, in an integrable model with a thermodynamic Bethe Ansatz description~\cite{Yang1969Thermodynamics,korepin_bogoliubov_izergin_1993}, filling functions characterise excited states in the thermodynamic limit and represent the fraction of particle excitations 
per given momentum~\cite{Mossel2012Generalized}; in noninteracting models $\vartheta(k)\sim \braket{\bs b^\dag_k \bs b^{\phantom \dag}_k}$ is the coarse grained fermion occupation number. For example, in the Fermi sea equivalent to the ground state of the XX model in zero field we have $\vartheta(k)=\frac{1+\mathrm{sgn}(\cos(k))}{2}$. 
We are not aware of theorems connecting  discontinuities in $\vartheta(k)$ with conformal invariance when $|2\vartheta(k)-1|\neq 1$,
but we stress that discontinuities
produce nevertheless algebraically decaying correlations.

For the sake of simplicity, we restrict ourselves to filling functions with symmetric discontinuities $\lim_{k\rightarrow k_{F}^\pm}(2\vartheta(k)-1)=\tanh(\eta_{k_{F}}^\pm)$, with $\eta_{k_{\textsc{f}}}^-=-\eta_{k_{\textsc{f}}}^+$.
We find that the large-distance properties of the state can be described by a quantum field theory of massless Dirac fermions, whose Euclidian action reads~\cite{Maric2023UniversalityLong}
\be\label{eq:S}
S=\int\mathrm d x\int\mathrm d \tau\sum_{k_{\textsc{f}}}\sum_{s=\pm}  \psi_{s,k_{\textsc{f}}}^\dag(\partial_{ \tau}+is v(k_{\textsc{f}})\partial_x) \psi_{s,k_{\textsc{f}}}\, ,
\ee  
where $k_{\textsc{f}}$ distinguishes theories emerging in the expansion around different discontinuities, $v(k_{\textsc{f}})$ is the velocity, and the fields satisfy standard anticommutation relations $\{ \psi_{s,k_{\textsc{f}}}(x), \psi^\dag_{s',k_{\textsc{f}}'}(y)\}=\delta_{s s'}\delta_{k_{\textsc{f}},k_{\textsc{f}}'}\delta(x-y)\mathrm I$, $\{ \psi_{s,k_{\textsc{f}}}(x), \psi_{s',k_{\textsc{f}}'}(y)\}=0$.
The standard procedure to compute the R\'enyi entropies of subsystems starts with reinterpreting the moments of the reduced density matrix, $\mathrm{tr}[\rho_X^\alpha]$, as partition functions (cf. Fig. 1 of Ref.~\cite{Calabrese2009Entanglement}) in models formed by $\alpha$ copies of the original one under 
the condition that each field is identified with the successive copy of itself when crossing the space-time lines corresponding to
$X$ at fixed (imaginary) time $\tau=0$. The moments are finally conveniently identified with the correlation functions of \emph{local} branch twist fields (associated with the global symmetry of exchange of the copies)~\cite{Cardy2008Form},  which are localised at the boundaries of $X$ and implicitly defined through the partition functions.

In spin chains, if the subsystem consists of two disjoint blocks, there is an additional complication related to the non-locality of the Jordan-Wigner transformation, indeed there are spin operators in $X$ whose fermionic representation includes fermions lying in between the blocks. This problem can be overcome by writing the reduced density matrix as a linear combination of four density matrices~\cite{Fagotti2010disjoint}. One should then generalise the field theory description so as to capture such a richer structure --- see, e.g., (Sec. 4 of) Ref.~\cite{Coser2016Spin}. For example, one  encounters also  terms of the form $\mathrm{tr}[\cdots P_A\rho_{A\cup C}^{n_j} P_{A}\rho_{A\cup C}^{n_{j+1}}\cdots]$, where $P_A$ counts the parity of the fermions in the first block $A$. In the QFT language, $P_A$ corresponds to the transformation that changes the sign of the field when crossing $A$.
In general there are also other contributions with the same root, but they are multiplied by expectation values of strings of Pauli matrices, which in our nonequilibrium setting decay exponentially with the separation. Accordingly, these terms can be dropped (see \cite{Maric2023UniversalityLong} for more details). 
This is why, despite the similarity with Fermi seas, condition $|\eta_{k_{\textsc{f}}}^\pm|\rightarrow\infty$ after global quenches generally leads to an unusual tripartite information: the value of the filling function at the discontinuities does not characterise the long-distance properties of all the relevant degrees of freedom, as a nontrivial behavior of the filling function between the discontinuities has the strong effect to turn some algebraically decaying correlations into exponentially decaying ones.

In our situation, in turn, a part of the structure behind the QFT procedure sketched above is lost. Specifically, the action of the $\alpha$-copy model is still simply the sum of $\alpha$ copies of $S$~\eqref{eq:S}, but only a ``physical part'' of the fields, namely,
$
 \Psi^{phys}_{k_{\textsc{f}}}(x)=\sum_s[1+\exp(2s \eta_{k_F}^+)]^{-1/2} \psi_{s,k_{\textsc{f}}}(x)
$
satisfies the standard conditions relating different copies on $X$ at zero imaginary time (whereas the ``unphysical parts" of the copies are independent).
And the same subtlety applies to $P_A$, which changes only the sign of the physical fields.  A direct consequence of having physical and unphysical fields  is that the energy density of the $\alpha$-copy model does change when crossing $X$ at $\tau=0$, undermining, e.g., the  interpretation of $\mathrm{tr}[\rho_X^\alpha]$ as the correlation function of \emph{local} branch twist fields~\cite{Cardy2008Form}. 
As detailed in Ref.~\cite{Maric2023UniversalityLong}, however, it is anyway possible to work out the asymptotic behaviour of the R\'enyi entropies of connected and disconnected spin blocks using the resolvent method, similarly to what was done in Refs~\cite{Casini2009Entanglement,Casini2009reduced_density,Fries2019,Blanco2022} for fermions.

\paragraph{Results.} We announce here the simplest results, as they are already sufficient to unveil the most striking feature of the tripartite information.
To start with, we focus on the limit $|\eta_{k_{\textsc{f}}}^\pm|\rightarrow\infty$. Since the same condition is satisfied in a Fermi sea, we can use the correspondence with the ground state of a CFT and take advantage of the results of Ref.~\cite{Coser2016Spin}, which computed the terms $\mathrm{tr}[\cdots P_A\rho_{A\cup C}^{n_j} P_{A}\rho_{A\cup C}^{n_{j+1}}\cdots]$ contributing to the tripartite information in the CFT ground state one by one (together with terms that do not contribute in our case).

The resulting prediction reads
\be\label{eq:Galphapred}
G_\alpha(x)\xrightarrow{|\eta_{k_{\textsc{f}}}^\pm|\rightarrow\infty}\frac{\log\Bigl[\sum_{\delta_j\in\{0,\frac{1}{2}\}\atop j=1,\dots,\alpha-1}\bigl(\tfrac{\Theta(\vec \delta|\hat\tau_x)}{\Theta(\vec 0|\hat\tau_x)}\bigr)^\nu\Bigr]}{\alpha-1}-\log 2
\ee
where $\hat \tau_x$ is the $(\alpha-1)\times (\alpha-1)$ period matrix of the Riemann surface $\mathcal R_\alpha$ with elements
\be
[\hat \tau_x]_{\ell n}=\frac{2i}{\alpha}\sum_{k=1}^{\alpha-1}\sin(\tfrac{\pi k}{\alpha})\cos(\tfrac{2\pi k(\ell-n)}{\alpha})\tfrac{P_{(k/\alpha)-1}(2x-1)}{P_{(k/\alpha)-1}(1-2x)}\, .
\ee
Here $P_\mu(z)$ denotes the Legendre functions, $
\Theta(\vec z,M)=\sum_{\vec m\in \mathbb Z^{\alpha-1}}e^{i\pi \vec m^t M \vec m+2\pi i \vec m\cdot \vec\delta}
$ is the Siegel theta function, 
and $\nu$ is the number of discontinuities of the filling function (assuming $|\eta_{k_{\textsc{f}}}^{\pm}|\rightarrow\infty$ for each one of them). 
The quench in Fig~\ref{f:2} can be used to check this prediction, indeed  $|\eta_{k_{\textsc{f}}}^\pm|\rightarrow\infty$. The agreement between numerical data and prediction is excellent. We stress that, contrary to the CFT analogue, our system does \emph{not} exhibit crossing symmetry, indeed  $G_\alpha(x)\neq G_\alpha(1-x)$.

Remarkably,  $\lim_{x\rightarrow 1^-}G_\alpha(x)=-\log 2$ for every $\alpha$, therefore (by the replica trick) we conclude that also the genuine tripartite information approaches the same value
\be\label{eq:G1}
\boxed{\mathfrak I_3=\lim_{x\rightarrow 1^-}G(x)=-\log 2}\, .
\ee
This limit corresponds to small separation between the blocks compared to their size, i.e. to the limit $1\ll|B|\ll |A|,|C|$. We call it \emph{residual tripartite information}, because $x$ is exactly equal to $1$ only when $|B|=0$, for which the tripartite information is zero by definition. Such unusual nonzero residual tripartite information should be contrasted to the ordinary zero value, which is found both in other nonequilibrium settings, such as after translationally invariant quenches from ground states of gapped Hamiltonians and critical Hamiltonians with $c=\frac{1}{2}$, and in equilibrium at any temperature, independently of criticality.

Equation~\eqref{eq:G1} is the main result of these notes. In all investigated systems with a nonzero tripartite information our analysis points to a \textit{universal} residual tripartite information equal to $-\log 2$, irrespectively of the quench protocol. 
In support of it, we also announce  $G_2(x)$ with generic $\eta_{k_{\textsc{f}}}^-$ and $\eta_{k_{\textsc{f}}}^+$:
\be
\label{eq:G2conj}
G_2(x)=\log\big(\tfrac{1+(1-x)^{\gamma}}{2}\big)\quad\Rightarrow G_2(1^-)=-\log 2\, ,
\ee
where $\gamma=\sum_{k_{\textsc{f}}}[\tfrac{1}{\pi}\arg(\sin(\frac{\pi}{4}+i\eta_{k_{\textsc{f}}}^+)/\sin(\frac{\pi}{4}+i\eta_{k_{\textsc{f}}}^-))]^2$.
Fig.~\ref{f:1} shows the excellent agreement of prediction\eqref{eq:G2conj} with numerical data. 

\paragraph{Numerical method.}
The R\'enyi entropies have been numerically evaluated  using their expressions in terms of the fermionic correlation matrix $\Gamma_{\ell n}=\delta_{\ell n}-\braket{\bs a_\ell \bs a_n}$. The latter has been computed directly in the generalised Gibbs ensemble emerging in the limit of infinite time~\cite{Jaynes1957,Rigol2007Relaxation}. 
For the entropy of connected subsystems we used  $S_\alpha(X)=\mathrm{tr}\bigl[\log\bigl((\frac{\mathrm I_X+\Gamma_X}{2})^\alpha+(\frac{\mathrm I_X-\Gamma_X}{2})^\alpha\bigr)\bigr]/[2(1-\alpha)]$, where $\Gamma_{X}$ is the correlation matrix in $X$~\cite{Vidal2003Entanglement,Jin2004}.  For the entropy of disjoint blocks in the generalised XY model we used the algorithm proposed in Ref.~\cite{Fagotti2010disjoint}, which allows one to express $S_\alpha(A\cup C)$ in terms of four matrices: $\Gamma_1\equiv \Gamma_{A\cup C}$, $\Gamma_2\equiv P_A \Gamma_1 P_A$, $\Gamma_3\equiv \Gamma_1-\Gamma_{A\cup C,B}\Gamma_{B}^{-1}\Gamma_{B,A\cup C}$, $\Gamma_4=P_A\Gamma_3 P_A$, where $\Gamma_{A,A'}$  is the correlation matrix in which the row and column indices run in $A$ and $A'$, respectively (we refer the reader to Sec. 3 of Ref.~\cite{Fagotti2010disjoint} for the formula with $\alpha=2,3,4$~\footnote{We warn the reader of a typo in Ref.~\cite{Fagotti2010disjoint}: $\det \Gamma_{B_1}$ in Eqs (57)-(59) should be replaced by $|\det \Gamma_{B_1}|$}). The terms claimed before to survive the infinite time limit are  those constructed with $\Gamma_1$ and $\Gamma_2$ only. 
For disjoint blocks in the dual XY model we have generalised the previous algorithm following Ref.~\cite{Fagotti2022Nonequilibrium}, and it is detailed elsewhere~\cite{Maric2023universality2}. 

\paragraph{Discussion.}
We have shown that the R\'enyi-$\alpha$ tripartite information captures universal properties in the limit of infinite time after a global quench whenever the correlations in the stationary state decay sufficiently slow with the distance. We have  provided evidence that $I_3^{(\alpha)}$ can be obtained within a quantum field theory and, in some cases, we have been able to predict their asymptotic behaviour. 
We think that the proposed framework could be generalised at least to interacting integrable systems that can be described by thermodynamic Bethe Ansatz, which have similar relaxation properties. If nonzero, we found always negative $I_3^{(\alpha)}$; we wonder how general in our setting this property is. Finally, we defined a residual tripartite information $\mathfrak I_3$, which we found either equal to $0$ or to $-\log 2$. We leave the question open of which values $\mathfrak I_3$ could attain with interactions. 

\begin{acknowledgments}
We are grateful to Leonardo Mazza for collaboration at an early stage of this project. We thank Raoul Santachiara for valuable discussions. We also thank Pasquale Calabrese and Erik Tonni for useful correspondence. 

This work was supported by the European Research Council under the Starting Grant No. 805252 LoCoMacro.
\end{acknowledgments}

\bibliography{references_published}

\begin{thebibliography}{58}%
\makeatletter
\providecommand \@ifxundefined [1]{%
 \@ifx{#1\undefined}
}%
\providecommand \@ifnum [1]{%
 \ifnum #1\expandafter \@firstoftwo
 \else \expandafter \@secondoftwo
 \fi
}%
\providecommand \@ifx [1]{%
 \ifx #1\expandafter \@firstoftwo
 \else \expandafter \@secondoftwo
 \fi
}%
\providecommand \natexlab [1]{#1}%
\providecommand \enquote  [1]{``#1''}%
\providecommand \bibnamefont  [1]{#1}%
\providecommand \bibfnamefont [1]{#1}%
\providecommand \citenamefont [1]{#1}%
\providecommand \href@noop [0]{\@secondoftwo}%
\providecommand \href [0]{\begingroup \@sanitize@url \@href}%
\providecommand \@href[1]{\@@startlink{#1}\@@href}%
\providecommand \@@href[1]{\endgroup#1\@@endlink}%
\providecommand \@sanitize@url [0]{\catcode `\\12\catcode `\$12\catcode
  `\&12\catcode `\#12\catcode `\^12\catcode `\_12\catcode `\%12\relax}%
\providecommand \@@startlink[1]{}%
\providecommand \@@endlink[0]{}%
\providecommand \url  [0]{\begingroup\@sanitize@url \@url }%
\providecommand \@url [1]{\endgroup\@href {#1}{\urlprefix }}%
\providecommand \urlprefix  [0]{URL }%
\providecommand \Eprint [0]{\href }%
\providecommand \doibase [0]{https://doi.org/}%
\providecommand \selectlanguage [0]{\@gobble}%
\providecommand \bibinfo  [0]{\@secondoftwo}%
\providecommand \bibfield  [0]{\@secondoftwo}%
\providecommand \translation [1]{[#1]}%
\providecommand \BibitemOpen [0]{}%
\providecommand \bibitemStop [0]{}%
\providecommand \bibitemNoStop [0]{.\EOS\space}%
\providecommand \EOS [0]{\spacefactor3000\relax}%
\providecommand \BibitemShut  [1]{\csname bibitem#1\endcsname}%
\let\auto@bib@innerbib\@empty
\bibitem [{\citenamefont {Einstein}\ \emph {et~al.}(1935)\citenamefont
  {Einstein}, \citenamefont {Podolsky},\ and\ \citenamefont
  {Rosen}}]{Einstein1935Can}%
  \BibitemOpen
  \bibfield  {author} {\bibinfo {author} {\bibfnamefont {A.}~\bibnamefont
  {Einstein}}, \bibinfo {author} {\bibfnamefont {B.}~\bibnamefont {Podolsky}},\
  and\ \bibinfo {author} {\bibfnamefont {N.}~\bibnamefont {Rosen}},\ }\bibfield
   {title} {\bibinfo {title} {{Can Quantum-Mechanical Description of Physical
  Reality Be Considered Complete?}},\ }\href
  {https://doi.org/10.1103/PhysRev.47.777} {\bibfield  {journal} {\bibinfo
  {journal} {Phys. Rev.}\ }\textbf {\bibinfo {volume} {47}},\ \bibinfo {pages}
  {777} (\bibinfo {year} {1935})}\BibitemShut {NoStop}%
\bibitem [{\citenamefont {Bell}(1964)}]{Bell1964On}%
  \BibitemOpen
  \bibfield  {author} {\bibinfo {author} {\bibfnamefont {J.~S.}\ \bibnamefont
  {Bell}},\ }\bibfield  {title} {\bibinfo {title} {{On the Einstein Podolsky
  Rosen paradox}},\ }\href
  {https://doi.org/10.1103/PhysicsPhysiqueFizika.1.195} {\bibfield  {journal}
  {\bibinfo  {journal} {Physics Physique Fizika}\ }\textbf {\bibinfo {volume}
  {1}},\ \bibinfo {pages} {195} (\bibinfo {year} {1964})}\BibitemShut {NoStop}%
\bibitem [{\citenamefont {Ingarden}(1976)}]{Ingarden1976Quantum}%
  \BibitemOpen
  \bibfield  {author} {\bibinfo {author} {\bibfnamefont {R.~S.}\ \bibnamefont
  {Ingarden}},\ }\bibfield  {title} {\bibinfo {title} {{Quantum information
  theory}},\ }\href
  {https://doi.org/https://doi.org/10.1016/0034-4877(76)90005-7} {\bibfield
  {journal} {\bibinfo  {journal} {Reports on Mathematical Physics}\ }\textbf
  {\bibinfo {volume} {10}},\ \bibinfo {pages} {43} (\bibinfo {year}
  {1976})}\BibitemShut {NoStop}%
\bibitem [{\citenamefont {Cerf}\ and\ \citenamefont
  {Adami}(1998)}]{Cerf1998Information}%
  \BibitemOpen
  \bibfield  {author} {\bibinfo {author} {\bibfnamefont {N.~J.}\ \bibnamefont
  {Cerf}}\ and\ \bibinfo {author} {\bibfnamefont {C.}~\bibnamefont {Adami}},\
  }\bibfield  {title} {\bibinfo {title} {{Information theory of quantum
  entanglement and measurement}},\ }\href
  {https://doi.org/https://doi.org/10.1016/S0167-2789(98)00045-1} {\bibfield
  {journal} {\bibinfo  {journal} {Physica D: Nonlinear Phenomena}\ }\textbf
  {\bibinfo {volume} {120}},\ \bibinfo {pages} {62} (\bibinfo {year} {1998})},\
  \bibinfo {note} {proceedings of the Fourth Workshop on Physics and
  Consumption}\BibitemShut {NoStop}%
\bibitem [{\citenamefont {Horodecki}\ \emph {et~al.}(2009)\citenamefont
  {Horodecki}, \citenamefont {Horodecki}, \citenamefont {Horodecki},\ and\
  \citenamefont {Horodecki}}]{Horodecki2009Quantum}%
  \BibitemOpen
  \bibfield  {author} {\bibinfo {author} {\bibfnamefont {R.}~\bibnamefont
  {Horodecki}}, \bibinfo {author} {\bibfnamefont {P.}~\bibnamefont
  {Horodecki}}, \bibinfo {author} {\bibfnamefont {M.}~\bibnamefont
  {Horodecki}},\ and\ \bibinfo {author} {\bibfnamefont {K.}~\bibnamefont
  {Horodecki}},\ }\bibfield  {title} {\bibinfo {title} {{Quantum
  entanglement}},\ }\href {https://doi.org/10.1103/RevModPhys.81.865}
  {\bibfield  {journal} {\bibinfo  {journal} {Rev. Mod. Phys.}\ }\textbf
  {\bibinfo {volume} {81}},\ \bibinfo {pages} {865} (\bibinfo {year}
  {2009})}\BibitemShut {NoStop}%
\bibitem [{\citenamefont {Holzhey}\ \emph {et~al.}(1994)\citenamefont
  {Holzhey}, \citenamefont {Larsen},\ and\ \citenamefont
  {Wilczek}}]{Holzhey1994Geometric}%
  \BibitemOpen
  \bibfield  {author} {\bibinfo {author} {\bibfnamefont {C.}~\bibnamefont
  {Holzhey}}, \bibinfo {author} {\bibfnamefont {F.}~\bibnamefont {Larsen}},\
  and\ \bibinfo {author} {\bibfnamefont {F.}~\bibnamefont {Wilczek}},\
  }\bibfield  {title} {\bibinfo {title} {Geometric and renormalized entropy in
  conformal field theory},\ }\href
  {https://doi.org/https://doi.org/10.1016/0550-3213(94)90402-2} {\bibfield
  {journal} {\bibinfo  {journal} {Nuclear Physics B}\ }\textbf {\bibinfo
  {volume} {424}},\ \bibinfo {pages} {443} (\bibinfo {year}
  {1994})}\BibitemShut {NoStop}%
\bibitem [{\citenamefont {Calabrese}\ and\ \citenamefont
  {Cardy}(2004)}]{Calabrese2004Entanglement}%
  \BibitemOpen
  \bibfield  {author} {\bibinfo {author} {\bibfnamefont {P.}~\bibnamefont
  {Calabrese}}\ and\ \bibinfo {author} {\bibfnamefont {J.}~\bibnamefont
  {Cardy}},\ }\bibfield  {title} {\bibinfo {title} {{Entanglement entropy and
  quantum field theory}},\ }\href
  {https://doi.org/10.1088/1742-5468/2004/06/p06002} {\bibfield  {journal}
  {\bibinfo  {journal} {Journal of Statistical Mechanics: Theory and
  Experiment}\ }\textbf {\bibinfo {volume} {2004}},\ \bibinfo {pages} {P06002}
  (\bibinfo {year} {2004})}\BibitemShut {NoStop}%
\bibitem [{\citenamefont {Korepin}(2004)}]{Korepin2004PRL}%
  \BibitemOpen
  \bibfield  {author} {\bibinfo {author} {\bibfnamefont {V.~E.}\ \bibnamefont
  {Korepin}},\ }\bibfield  {title} {\bibinfo {title} {Universality of entropy
  scaling in one dimensional gapless models},\ }\href
  {https://doi.org/10.1103/PhysRevLett.92.096402} {\bibfield  {journal}
  {\bibinfo  {journal} {Phys. Rev. Lett.}\ }\textbf {\bibinfo {volume} {92}},\
  \bibinfo {pages} {096402} (\bibinfo {year} {2004})}\BibitemShut {NoStop}%
\bibitem [{\citenamefont {Caraglio}\ and\ \citenamefont
  {Gliozzi}(2008)}]{Caraglio2008Entanglement}%
  \BibitemOpen
  \bibfield  {author} {\bibinfo {author} {\bibfnamefont {M.}~\bibnamefont
  {Caraglio}}\ and\ \bibinfo {author} {\bibfnamefont {F.}~\bibnamefont
  {Gliozzi}},\ }\bibfield  {title} {\bibinfo {title} {{Entanglement entropy and
  twist fields}},\ }\href {https://doi.org/10.1088/1126-6708/2008/11/076}
  {\bibfield  {journal} {\bibinfo  {journal} {Journal of High Energy Physics}\
  }\textbf {\bibinfo {volume} {2008}},\ \bibinfo {pages} {076} (\bibinfo {year}
  {2008})}\BibitemShut {NoStop}%
\bibitem [{\citenamefont {Furukawa}\ \emph {et~al.}(2009)\citenamefont
  {Furukawa}, \citenamefont {Pasquier},\ and\ \citenamefont
  {Shiraishi}}]{Furukawa2009Mutual}%
  \BibitemOpen
  \bibfield  {author} {\bibinfo {author} {\bibfnamefont {S.}~\bibnamefont
  {Furukawa}}, \bibinfo {author} {\bibfnamefont {V.}~\bibnamefont {Pasquier}},\
  and\ \bibinfo {author} {\bibfnamefont {J.}~\bibnamefont {Shiraishi}},\
  }\bibfield  {title} {\bibinfo {title} {{Mutual Information and Boson Radius
  in a $c=1$ Critical System in One Dimension}},\ }\href
  {https://doi.org/10.1103/PhysRevLett.102.170602} {\bibfield  {journal}
  {\bibinfo  {journal} {Phys. Rev. Lett.}\ }\textbf {\bibinfo {volume} {102}},\
  \bibinfo {pages} {170602} (\bibinfo {year} {2009})}\BibitemShut {NoStop}%
\bibitem [{\citenamefont {Calabrese}\ and\ \citenamefont
  {Cardy}(2009)}]{Calabrese2009Entanglement}%
  \BibitemOpen
  \bibfield  {author} {\bibinfo {author} {\bibfnamefont {P.}~\bibnamefont
  {Calabrese}}\ and\ \bibinfo {author} {\bibfnamefont {J.}~\bibnamefont
  {Cardy}},\ }\bibfield  {title} {\bibinfo {title} {{Entanglement entropy and
  conformal field theory}},\ }\href
  {https://doi.org/10.1088/1751-8113/42/50/504005} {\bibfield  {journal}
  {\bibinfo  {journal} {Journal of Physics A: Mathematical and Theoretical}\
  }\textbf {\bibinfo {volume} {42}},\ \bibinfo {pages} {504005} (\bibinfo
  {year} {2009})}\BibitemShut {NoStop}%
\bibitem [{\citenamefont {Kitaev}\ and\ \citenamefont
  {Preskill}(2006)}]{Kitaev2006Topological}%
  \BibitemOpen
  \bibfield  {author} {\bibinfo {author} {\bibfnamefont {A.}~\bibnamefont
  {Kitaev}}\ and\ \bibinfo {author} {\bibfnamefont {J.}~\bibnamefont
  {Preskill}},\ }\bibfield  {title} {\bibinfo {title} {{Topological
  Entanglement Entropy}},\ }\href
  {https://doi.org/10.1103/PhysRevLett.96.110404} {\bibfield  {journal}
  {\bibinfo  {journal} {Phys. Rev. Lett.}\ }\textbf {\bibinfo {volume} {96}},\
  \bibinfo {pages} {110404} (\bibinfo {year} {2006})}\BibitemShut {NoStop}%
\bibitem [{\citenamefont {Casini}\ and\ \citenamefont
  {Huerta}(2009{\natexlab{a}})}]{Casini2009Remarks}%
  \BibitemOpen
  \bibfield  {author} {\bibinfo {author} {\bibfnamefont {H.}~\bibnamefont
  {Casini}}\ and\ \bibinfo {author} {\bibfnamefont {M.}~\bibnamefont
  {Huerta}},\ }\bibfield  {title} {\bibinfo {title} {{Remarks on the
  entanglement entropy for disconnected regions}},\ }\href
  {https://doi.org/10.1088/1126-6708/2009/03/048} {\bibfield  {journal}
  {\bibinfo  {journal} {Journal of High Energy Physics}\ }\textbf {\bibinfo
  {volume} {2009}},\ \bibinfo {pages} {048} (\bibinfo {year}
  {2009}{\natexlab{a}})}\BibitemShut {NoStop}%
\bibitem [{\citenamefont {Ag\'on}\ \emph {et~al.}(2022)\citenamefont {Ag\'on},
  \citenamefont {Bueno},\ and\ \citenamefont {Casini}}]{Agon2022Tripartite}%
  \BibitemOpen
  \bibfield  {author} {\bibinfo {author} {\bibfnamefont {C.~A.}\ \bibnamefont
  {Ag\'on}}, \bibinfo {author} {\bibfnamefont {P.}~\bibnamefont {Bueno}},\ and\
  \bibinfo {author} {\bibfnamefont {H.}~\bibnamefont {Casini}},\ }\bibfield
  {title} {\bibinfo {title} {{Tripartite information at long distances}},\
  }\href {https://doi.org/10.21468/SciPostPhys.12.5.153} {\bibfield  {journal}
  {\bibinfo  {journal} {SciPost Phys.}\ }\textbf {\bibinfo {volume} {12}},\
  \bibinfo {pages} {153} (\bibinfo {year} {2022})}\BibitemShut {NoStop}%
\bibitem [{\citenamefont {Hayden}\ \emph {et~al.}(2013)\citenamefont {Hayden},
  \citenamefont {Headrick},\ and\ \citenamefont
  {Maloney}}]{Hayden2013Holographic}%
  \BibitemOpen
  \bibfield  {author} {\bibinfo {author} {\bibfnamefont {P.}~\bibnamefont
  {Hayden}}, \bibinfo {author} {\bibfnamefont {M.}~\bibnamefont {Headrick}},\
  and\ \bibinfo {author} {\bibfnamefont {A.}~\bibnamefont {Maloney}},\
  }\bibfield  {title} {\bibinfo {title} {Holographic mutual information is
  monogamous},\ }\href {https://doi.org/10.1103/PhysRevD.87.046003} {\bibfield
  {journal} {\bibinfo  {journal} {Phys. Rev. D}\ }\textbf {\bibinfo {volume}
  {87}},\ \bibinfo {pages} {046003} (\bibinfo {year} {2013})}\BibitemShut
  {NoStop}%
\bibitem [{\citenamefont {Hosur}\ \emph {et~al.}(2016)\citenamefont {Hosur},
  \citenamefont {Qi}, \citenamefont {Roberts},\ and\ \citenamefont
  {Yoshida}}]{Hosur2016Chaos}%
  \BibitemOpen
  \bibfield  {author} {\bibinfo {author} {\bibfnamefont {P.}~\bibnamefont
  {Hosur}}, \bibinfo {author} {\bibfnamefont {X.-L.}\ \bibnamefont {Qi}},
  \bibinfo {author} {\bibfnamefont {D.~A.}\ \bibnamefont {Roberts}},\ and\
  \bibinfo {author} {\bibfnamefont {B.}~\bibnamefont {Yoshida}},\ }\bibfield
  {title} {\bibinfo {title} {{Chaos in quantum channels}},\ }\href
  {https://doi.org/10.1007/JHEP02(2016)004} {\bibfield  {journal} {\bibinfo
  {journal} {Journal of High Energy Physics}\ }\textbf {\bibinfo {volume}
  {2016}},\ \bibinfo {pages} {4} (\bibinfo {year} {2016})}\BibitemShut
  {NoStop}%
\bibitem [{\citenamefont {Schnaack}\ \emph {et~al.}(2019)\citenamefont
  {Schnaack}, \citenamefont {B\"olter}, \citenamefont {Paeckel}, \citenamefont
  {Manmana}, \citenamefont {Kehrein},\ and\ \citenamefont
  {Schmitt}}]{Schnaack2019Tripartite}%
  \BibitemOpen
  \bibfield  {author} {\bibinfo {author} {\bibfnamefont {O.}~\bibnamefont
  {Schnaack}}, \bibinfo {author} {\bibfnamefont {N.}~\bibnamefont {B\"olter}},
  \bibinfo {author} {\bibfnamefont {S.}~\bibnamefont {Paeckel}}, \bibinfo
  {author} {\bibfnamefont {S.~R.}\ \bibnamefont {Manmana}}, \bibinfo {author}
  {\bibfnamefont {S.}~\bibnamefont {Kehrein}},\ and\ \bibinfo {author}
  {\bibfnamefont {M.}~\bibnamefont {Schmitt}},\ }\bibfield  {title} {\bibinfo
  {title} {{Tripartite information, scrambling, and the role of Hilbert space
  partitioning in quantum lattice models}},\ }\href
  {https://doi.org/10.1103/PhysRevB.100.224302} {\bibfield  {journal} {\bibinfo
   {journal} {Phys. Rev. B}\ }\textbf {\bibinfo {volume} {100}},\ \bibinfo
  {pages} {224302} (\bibinfo {year} {2019})}\BibitemShut {NoStop}%
\bibitem [{\citenamefont {S{\"u}nderhauf}\ \emph {et~al.}(2019)\citenamefont
  {S{\"u}nderhauf}, \citenamefont {Piroli}, \citenamefont {Qi}, \citenamefont
  {Schuch},\ and\ \citenamefont {Cirac}}]{Sunderhauf2019Quantum}%
  \BibitemOpen
  \bibfield  {author} {\bibinfo {author} {\bibfnamefont {C.}~\bibnamefont
  {S{\"u}nderhauf}}, \bibinfo {author} {\bibfnamefont {L.}~\bibnamefont
  {Piroli}}, \bibinfo {author} {\bibfnamefont {X.-L.}\ \bibnamefont {Qi}},
  \bibinfo {author} {\bibfnamefont {N.}~\bibnamefont {Schuch}},\ and\ \bibinfo
  {author} {\bibfnamefont {J.~I.}\ \bibnamefont {Cirac}},\ }\bibfield  {title}
  {\bibinfo {title} {{Quantum chaos in the Brownian SYK model with large finite
  N : OTOCs and tripartite information}},\ }\href
  {https://doi.org/10.1007/JHEP11(2019)038} {\bibfield  {journal} {\bibinfo
  {journal} {Journal of High Energy Physics}\ }\textbf {\bibinfo {volume}
  {2019}},\ \bibinfo {pages} {38} (\bibinfo {year} {2019})}\BibitemShut
  {NoStop}%
\bibitem [{\citenamefont {Coser}\ \emph
  {et~al.}(2014{\natexlab{a}})\citenamefont {Coser}, \citenamefont
  {Tagliacozzo},\ and\ \citenamefont {Tonni}}]{Coser2014On}%
  \BibitemOpen
  \bibfield  {author} {\bibinfo {author} {\bibfnamefont {A.}~\bibnamefont
  {Coser}}, \bibinfo {author} {\bibfnamefont {L.}~\bibnamefont {Tagliacozzo}},\
  and\ \bibinfo {author} {\bibfnamefont {E.}~\bibnamefont {Tonni}},\ }\bibfield
   {title} {\bibinfo {title} {{On R{\'{e}}nyi entropies of disjoint intervals
  in conformal field theory}},\ }\href
  {https://doi.org/10.1088/1742-5468/2014/01/p01008} {\bibfield  {journal}
  {\bibinfo  {journal} {Journal of Statistical Mechanics: Theory and
  Experiment}\ }\textbf {\bibinfo {volume} {2014}},\ \bibinfo {pages} {P01008}
  (\bibinfo {year} {2014}{\natexlab{a}})}\BibitemShut {NoStop}%
\bibitem [{\citenamefont {Nobili}\ \emph {et~al.}(2015)\citenamefont {Nobili},
  \citenamefont {Coser},\ and\ \citenamefont
  {Tonni}}]{DeNobili2015Entanglement}%
  \BibitemOpen
  \bibfield  {author} {\bibinfo {author} {\bibfnamefont {C.~D.}\ \bibnamefont
  {Nobili}}, \bibinfo {author} {\bibfnamefont {A.}~\bibnamefont {Coser}},\ and\
  \bibinfo {author} {\bibfnamefont {E.}~\bibnamefont {Tonni}},\ }\bibfield
  {title} {\bibinfo {title} {{Entanglement entropy and negativity of disjoint
  intervals in {CFT}: some numerical extrapolations}},\ }\href
  {https://doi.org/10.1088/1742-5468/2015/06/p06021} {\bibfield  {journal}
  {\bibinfo  {journal} {Journal of Statistical Mechanics: Theory and
  Experiment}\ }\textbf {\bibinfo {volume} {2015}},\ \bibinfo {pages} {P06021}
  (\bibinfo {year} {2015})}\BibitemShut {NoStop}%
\bibitem [{\citenamefont {Calabrese}\ \emph {et~al.}(2009)\citenamefont
  {Calabrese}, \citenamefont {Cardy},\ and\ \citenamefont
  {Tonni}}]{Calabrese2009Entanglement1}%
  \BibitemOpen
  \bibfield  {author} {\bibinfo {author} {\bibfnamefont {P.}~\bibnamefont
  {Calabrese}}, \bibinfo {author} {\bibfnamefont {J.}~\bibnamefont {Cardy}},\
  and\ \bibinfo {author} {\bibfnamefont {E.}~\bibnamefont {Tonni}},\ }\bibfield
   {title} {\bibinfo {title} {{Entanglement entropy of two disjoint intervals
  in conformal field theory}},\ }\href
  {https://doi.org/10.1088/1742-5468/2009/11/P11001} {\bibfield  {journal}
  {\bibinfo  {journal} {Journal of Statistical Mechanics: Theory and
  Experiment}\ }\textbf {\bibinfo {volume} {2009}},\ \bibinfo {pages} {P11001}
  (\bibinfo {year} {2009})}\BibitemShut {NoStop}%
\bibitem [{\citenamefont {Calabrese}\ \emph {et~al.}(2011)\citenamefont
  {Calabrese}, \citenamefont {Cardy},\ and\ \citenamefont
  {Tonni}}]{Calabrese2011Entanglement}%
  \BibitemOpen
  \bibfield  {author} {\bibinfo {author} {\bibfnamefont {P.}~\bibnamefont
  {Calabrese}}, \bibinfo {author} {\bibfnamefont {J.}~\bibnamefont {Cardy}},\
  and\ \bibinfo {author} {\bibfnamefont {E.}~\bibnamefont {Tonni}},\ }\bibfield
   {title} {\bibinfo {title} {{Entanglement entropy of two disjoint intervals
  in conformal field theory: {II}}},\ }\href
  {https://doi.org/10.1088/1742-5468/2011/01/p01021} {\bibfield  {journal}
  {\bibinfo  {journal} {Journal of Statistical Mechanics: Theory and
  Experiment}\ }\textbf {\bibinfo {volume} {2011}},\ \bibinfo {pages} {P01021}
  (\bibinfo {year} {2011})}\BibitemShut {NoStop}%
\bibitem [{\citenamefont {Ares}\ \emph {et~al.}(2021)\citenamefont {Ares},
  \citenamefont {Santachiara},\ and\ \citenamefont {Viti}}]{Ares2021Crossing}%
  \BibitemOpen
  \bibfield  {author} {\bibinfo {author} {\bibfnamefont {F.}~\bibnamefont
  {Ares}}, \bibinfo {author} {\bibfnamefont {R.}~\bibnamefont {Santachiara}},\
  and\ \bibinfo {author} {\bibfnamefont {J.}~\bibnamefont {Viti}},\ }\bibfield
  {title} {\bibinfo {title} {{Crossing-symmetric twist field correlators and
  entanglement negativity in minimal CFTs}},\ }\href
  {https://doi.org/10.1007/JHEP10(2021)175} {\bibfield  {journal} {\bibinfo
  {journal} {Journal of High Energy Physics}\ }\textbf {\bibinfo {volume}
  {2021}},\ \bibinfo {pages} {175} (\bibinfo {year} {2021})}\BibitemShut
  {NoStop}%
\bibitem [{\citenamefont {Fagotti}\ and\ \citenamefont
  {Calabrese}(2010)}]{Fagotti2010disjoint}%
  \BibitemOpen
  \bibfield  {author} {\bibinfo {author} {\bibfnamefont {M.}~\bibnamefont
  {Fagotti}}\ and\ \bibinfo {author} {\bibfnamefont {P.}~\bibnamefont
  {Calabrese}},\ }\bibfield  {title} {\bibinfo {title} {{Entanglement entropy
  of two disjoint blocks in XY chains}},\ }\href
  {https://doi.org/10.1088/1742-5468/2010/04/p04016} {\bibfield  {journal}
  {\bibinfo  {journal} {Journal of Statistical Mechanics: Theory and
  Experiment}\ }\textbf {\bibinfo {volume} {2010}},\ \bibinfo {pages} {P04016}
  (\bibinfo {year} {2010})}\BibitemShut {NoStop}%
\bibitem [{\citenamefont {Alba}\ \emph {et~al.}(2010)\citenamefont {Alba},
  \citenamefont {Tagliacozzo},\ and\ \citenamefont
  {Calabrese}}]{Alba2010Entanglement}%
  \BibitemOpen
  \bibfield  {author} {\bibinfo {author} {\bibfnamefont {V.}~\bibnamefont
  {Alba}}, \bibinfo {author} {\bibfnamefont {L.}~\bibnamefont {Tagliacozzo}},\
  and\ \bibinfo {author} {\bibfnamefont {P.}~\bibnamefont {Calabrese}},\
  }\bibfield  {title} {\bibinfo {title} {Entanglement entropy of two disjoint
  blocks in critical ising models},\ }\href
  {https://doi.org/10.1103/PhysRevB.81.060411} {\bibfield  {journal} {\bibinfo
  {journal} {Phys. Rev. B}\ }\textbf {\bibinfo {volume} {81}},\ \bibinfo
  {pages} {060411} (\bibinfo {year} {2010})}\BibitemShut {NoStop}%
\bibitem [{\citenamefont {Alba}\ \emph {et~al.}(2011)\citenamefont {Alba},
  \citenamefont {Tagliacozzo},\ and\ \citenamefont
  {Calabrese}}]{Alba2011Entanglement}%
  \BibitemOpen
  \bibfield  {author} {\bibinfo {author} {\bibfnamefont {V.}~\bibnamefont
  {Alba}}, \bibinfo {author} {\bibfnamefont {L.}~\bibnamefont {Tagliacozzo}},\
  and\ \bibinfo {author} {\bibfnamefont {P.}~\bibnamefont {Calabrese}},\
  }\bibfield  {title} {\bibinfo {title} {{Entanglement entropy of two disjoint
  intervals in c=1 theories}},\ }\href
  {https://doi.org/10.1088/1742-5468/2011/06/p06012} {\bibfield  {journal}
  {\bibinfo  {journal} {Journal of Statistical Mechanics: Theory and
  Experiment}\ }\textbf {\bibinfo {volume} {2011}},\ \bibinfo {pages} {P06012}
  (\bibinfo {year} {2011})}\BibitemShut {NoStop}%
\bibitem [{\citenamefont {Fagotti}(2012)}]{Fagotti2012New}%
  \BibitemOpen
  \bibfield  {author} {\bibinfo {author} {\bibfnamefont {M.}~\bibnamefont
  {Fagotti}},\ }\bibfield  {title} {\bibinfo {title} {New insights into the
  entanglement of disjoint blocks},\ }\href
  {https://doi.org/10.1209/0295-5075/97/17007} {\bibfield  {journal} {\bibinfo
  {journal} {{EPL} (Europhysics Letters)}\ }\textbf {\bibinfo {volume} {97}},\
  \bibinfo {pages} {17007} (\bibinfo {year} {2012})}\BibitemShut {NoStop}%
\bibitem [{\citenamefont {Coser}\ \emph {et~al.}(2016)\citenamefont {Coser},
  \citenamefont {Tonni},\ and\ \citenamefont {Calabrese}}]{Coser2016Spin}%
  \BibitemOpen
  \bibfield  {author} {\bibinfo {author} {\bibfnamefont {A.}~\bibnamefont
  {Coser}}, \bibinfo {author} {\bibfnamefont {E.}~\bibnamefont {Tonni}},\ and\
  \bibinfo {author} {\bibfnamefont {P.}~\bibnamefont {Calabrese}},\ }\bibfield
  {title} {\bibinfo {title} {{Spin structures and entanglement of two disjoint
  intervals in conformal field theories}},\ }\href
  {https://doi.org/10.1088/1742-5468/2016/05/053109} {\bibfield  {journal}
  {\bibinfo  {journal} {Journal of Statistical Mechanics: Theory and
  Experiment}\ }\textbf {\bibinfo {volume} {2016}},\ \bibinfo {pages} {053109}
  (\bibinfo {year} {2016})}\BibitemShut {NoStop}%
\bibitem [{\citenamefont {Alba}\ and\ \citenamefont
  {Calabrese}(2017)}]{Alba2017Entanglement}%
  \BibitemOpen
  \bibfield  {author} {\bibinfo {author} {\bibfnamefont {V.}~\bibnamefont
  {Alba}}\ and\ \bibinfo {author} {\bibfnamefont {P.}~\bibnamefont
  {Calabrese}},\ }\bibfield  {title} {\bibinfo {title} {{Entanglement and
  thermodynamics after a quantum quench in integrable systems}},\ }\href
  {https://doi.org/10.1073/pnas.1703516114} {\bibfield  {journal} {\bibinfo
  {journal} {Proceedings of the National Academy of Sciences}\ }\textbf
  {\bibinfo {volume} {114}},\ \bibinfo {pages} {7947} (\bibinfo {year}
  {2017})},\ \Eprint
  {https://arxiv.org/abs/https://www.pnas.org/doi/pdf/10.1073/pnas.1703516114}
  {https://www.pnas.org/doi/pdf/10.1073/pnas.1703516114} \BibitemShut {NoStop}%
\bibitem [{\citenamefont {Bertini}\ \emph {et~al.}(2022)\citenamefont
  {Bertini}, \citenamefont {Klobas}, \citenamefont {Alba}, \citenamefont
  {Lagnese},\ and\ \citenamefont {Calabrese}}]{Bertini2022Growth}%
  \BibitemOpen
  \bibfield  {author} {\bibinfo {author} {\bibfnamefont {B.}~\bibnamefont
  {Bertini}}, \bibinfo {author} {\bibfnamefont {K.}~\bibnamefont {Klobas}},
  \bibinfo {author} {\bibfnamefont {V.}~\bibnamefont {Alba}}, \bibinfo {author}
  {\bibfnamefont {G.}~\bibnamefont {Lagnese}},\ and\ \bibinfo {author}
  {\bibfnamefont {P.}~\bibnamefont {Calabrese}},\ }\bibfield  {title} {\bibinfo
  {title} {Growth of r\'enyi entropies in interacting integrable models and the
  breakdown of the quasiparticle picture},\ }\href
  {https://doi.org/10.1103/PhysRevX.12.031016} {\bibfield  {journal} {\bibinfo
  {journal} {Phys. Rev. X}\ }\textbf {\bibinfo {volume} {12}},\ \bibinfo
  {pages} {031016} (\bibinfo {year} {2022})}\BibitemShut {NoStop}%
\bibitem [{\citenamefont {Zhou}\ and\ \citenamefont
  {Nahum}(2020)}]{Zhou2020Entanglement}%
  \BibitemOpen
  \bibfield  {author} {\bibinfo {author} {\bibfnamefont {T.}~\bibnamefont
  {Zhou}}\ and\ \bibinfo {author} {\bibfnamefont {A.}~\bibnamefont {Nahum}},\
  }\bibfield  {title} {\bibinfo {title} {Entanglement membrane in chaotic
  many-body systems},\ }\href {https://doi.org/10.1103/PhysRevX.10.031066}
  {\bibfield  {journal} {\bibinfo  {journal} {Phys. Rev. X}\ }\textbf {\bibinfo
  {volume} {10}},\ \bibinfo {pages} {031066} (\bibinfo {year}
  {2020})}\BibitemShut {NoStop}%
\bibitem [{\citenamefont {Srednicki}(1994)}]{srednicki94}%
  \BibitemOpen
  \bibfield  {author} {\bibinfo {author} {\bibfnamefont {M.}~\bibnamefont
  {Srednicki}},\ }\bibfield  {title} {\bibinfo {title} {{Chaos and quantum
  thermalization}},\ }\href {https://doi.org/10.1103/PhysRevE.50.888}
  {\bibfield  {journal} {\bibinfo  {journal} {Phys. Rev. E}\ }\textbf {\bibinfo
  {volume} {50}},\ \bibinfo {pages} {888} (\bibinfo {year} {1994})}\BibitemShut
  {NoStop}%
\bibitem [{\citenamefont {Polkovnikov}\ \emph {et~al.}(2011)\citenamefont
  {Polkovnikov}, \citenamefont {Sengupta}, \citenamefont {Silva},\ and\
  \citenamefont {Vengalattore}}]{Polkovnikov2011Colloquium}%
  \BibitemOpen
  \bibfield  {author} {\bibinfo {author} {\bibfnamefont {A.}~\bibnamefont
  {Polkovnikov}}, \bibinfo {author} {\bibfnamefont {K.}~\bibnamefont
  {Sengupta}}, \bibinfo {author} {\bibfnamefont {A.}~\bibnamefont {Silva}},\
  and\ \bibinfo {author} {\bibfnamefont {M.}~\bibnamefont {Vengalattore}},\
  }\bibfield  {title} {\bibinfo {title} {{Colloquium: Nonequilibrium dynamics
  of closed interacting quantum systems}},\ }\href
  {https://doi.org/10.1103/RevModPhys.83.863} {\bibfield  {journal} {\bibinfo
  {journal} {Rev. Mod. Phys.}\ }\textbf {\bibinfo {volume} {83}},\ \bibinfo
  {pages} {863} (\bibinfo {year} {2011})}\BibitemShut {NoStop}%
\bibitem [{\citenamefont {Gogolin}\ and\ \citenamefont
  {Eisert}(2016)}]{Gogolin2016Equilibration}%
  \BibitemOpen
  \bibfield  {author} {\bibinfo {author} {\bibfnamefont {C.}~\bibnamefont
  {Gogolin}}\ and\ \bibinfo {author} {\bibfnamefont {J.}~\bibnamefont
  {Eisert}},\ }\bibfield  {title} {\bibinfo {title} {{Equilibration,
  thermalisation, and the emergence of statistical mechanics in closed quantum
  systems}},\ }\href {https://doi.org/10.1088/0034-4885/79/5/056001} {\bibfield
   {journal} {\bibinfo  {journal} {Rep. Prog. Phys.}\ }\textbf {\bibinfo
  {volume} {79}},\ \bibinfo {pages} {056001} (\bibinfo {year}
  {2016})}\BibitemShut {NoStop}%
\bibitem [{\citenamefont {Suzuki}(1971)}]{Suzuki1971The}%
  \BibitemOpen
  \bibfield  {author} {\bibinfo {author} {\bibfnamefont {M.}~\bibnamefont
  {Suzuki}},\ }\bibfield  {title} {\bibinfo {title} {{The dimer problem and the
  generalized X-model}},\ }\href
  {https://doi.org/https://doi.org/10.1016/0375-9601(71)90901-7} {\bibfield
  {journal} {\bibinfo  {journal} {Physics Letters A}\ }\textbf {\bibinfo
  {volume} {34}},\ \bibinfo {pages} {338} (\bibinfo {year} {1971})}\BibitemShut
  {NoStop}%
\bibitem [{\citenamefont {Lieb}\ \emph {et~al.}(1961)\citenamefont {Lieb},
  \citenamefont {Schultz},\ and\ \citenamefont {Mattis}}]{Lieb1961}%
  \BibitemOpen
  \bibfield  {author} {\bibinfo {author} {\bibfnamefont {E.}~\bibnamefont
  {Lieb}}, \bibinfo {author} {\bibfnamefont {T.}~\bibnamefont {Schultz}},\ and\
  \bibinfo {author} {\bibfnamefont {D.}~\bibnamefont {Mattis}},\ }\bibfield
  {title} {\bibinfo {title} {{Two soluble models of an antiferromagnetic
  chain}},\ }\href
  {https://doi.org/https://doi.org/10.1016/0003-4916(61)90115-4} {\bibfield
  {journal} {\bibinfo  {journal} {Annals of Physics}\ }\textbf {\bibinfo
  {volume} {16}},\ \bibinfo {pages} {407} (\bibinfo {year} {1961})}\BibitemShut
  {NoStop}%
\bibitem [{\citenamefont {Zadnik}\ and\ \citenamefont
  {Fagotti}(2021)}]{Zadnik2021The}%
  \BibitemOpen
  \bibfield  {author} {\bibinfo {author} {\bibfnamefont {L.}~\bibnamefont
  {Zadnik}}\ and\ \bibinfo {author} {\bibfnamefont {M.}~\bibnamefont
  {Fagotti}},\ }\bibfield  {title} {\bibinfo {title} {{The Folded Spin-1/2 XXZ
  Model: I. Diagonalisation, Jamming, and Ground State Properties}},\ }\href
  {https://doi.org/10.21468/SciPostPhysCore.4.2.010} {\bibfield  {journal}
  {\bibinfo  {journal} {SciPost Phys. Core}\ }\textbf {\bibinfo {volume} {4}},\
  \bibinfo {pages} {10} (\bibinfo {year} {2021})}\BibitemShut {NoStop}%
\bibitem [{\citenamefont {Fagotti}(2022)}]{Fagotti2022Global}%
  \BibitemOpen
  \bibfield  {author} {\bibinfo {author} {\bibfnamefont {M.}~\bibnamefont
  {Fagotti}},\ }\bibfield  {title} {\bibinfo {title} {{Global Quenches after
  Localized Perturbations}},\ }\href
  {https://doi.org/10.1103/PhysRevLett.128.110602} {\bibfield  {journal}
  {\bibinfo  {journal} {Phys. Rev. Lett.}\ }\textbf {\bibinfo {volume} {128}},\
  \bibinfo {pages} {110602} (\bibinfo {year} {2022})}\BibitemShut {NoStop}%
\bibitem [{\citenamefont {Fagotti}\ \emph {et~al.}(2022)\citenamefont
  {Fagotti}, \citenamefont {Marić},\ and\ \citenamefont
  {Zadnik}}]{Fagotti2022Nonequilibrium}%
  \BibitemOpen
  \bibfield  {author} {\bibinfo {author} {\bibfnamefont {M.}~\bibnamefont
  {Fagotti}}, \bibinfo {author} {\bibfnamefont {V.}~\bibnamefont {Marić}},\
  and\ \bibinfo {author} {\bibfnamefont {L.}~\bibnamefont {Zadnik}},\
  }\bibfield  {title} {\bibinfo {title} {{Nonequilibrium symmetry-protected
  topological order: emergence of semilocal Gibbs ensembles}},\ }\Eprint
  {https://arxiv.org/abs/2205.02221} {arXiv:2205.02221}  (\bibinfo {year}
  {2022})\BibitemShut {NoStop}%
\bibitem [{\citenamefont {Essler}\ and\ \citenamefont
  {Fagotti}(2016)}]{Essler_2016}%
  \BibitemOpen
  \bibfield  {author} {\bibinfo {author} {\bibfnamefont {F.~H.~L.}\
  \bibnamefont {Essler}}\ and\ \bibinfo {author} {\bibfnamefont
  {M.}~\bibnamefont {Fagotti}},\ }\bibfield  {title} {\bibinfo {title} {{Quench
  dynamics and relaxation in isolated integrable quantum spin chains}},\ }\href
  {https://doi.org/10.1088/1742-5468/2016/06/064002} {\bibfield  {journal}
  {\bibinfo  {journal} {Journal of Statistical Mechanics: Theory and
  Experiment}\ }\textbf {\bibinfo {volume} {2016}},\ \bibinfo {pages} {064002}
  (\bibinfo {year} {2016})}\BibitemShut {NoStop}%
\bibitem [{\citenamefont {Coser}\ \emph
  {et~al.}(2014{\natexlab{b}})\citenamefont {Coser}, \citenamefont {Tonni},\
  and\ \citenamefont {Calabrese}}]{Coser2014Entanglement}%
  \BibitemOpen
  \bibfield  {author} {\bibinfo {author} {\bibfnamefont {A.}~\bibnamefont
  {Coser}}, \bibinfo {author} {\bibfnamefont {E.}~\bibnamefont {Tonni}},\ and\
  \bibinfo {author} {\bibfnamefont {P.}~\bibnamefont {Calabrese}},\ }\bibfield
  {title} {\bibinfo {title} {Entanglement negativity after a global quantum
  quench},\ }\href {https://doi.org/10.1088/1742-5468/2014/12/p12017}
  {\bibfield  {journal} {\bibinfo  {journal} {Journal of Statistical Mechanics:
  Theory and Experiment}\ }\textbf {\bibinfo {volume} {2014}},\ \bibinfo
  {pages} {P12017} (\bibinfo {year} {2014}{\natexlab{b}})}\BibitemShut
  {NoStop}%
\bibitem [{\citenamefont {Mari{\'{c}}}\ and\ \citenamefont
  {Fagotti}(2023)}]{Maric2023UniversalityLong}%
  \BibitemOpen
  \bibfield  {author} {\bibinfo {author} {\bibfnamefont {V.}~\bibnamefont
  {Mari{\'{c}}}}\ and\ \bibinfo {author} {\bibfnamefont {M.}~\bibnamefont
  {Fagotti}},\ }\bibfield  {title} {\bibinfo {title} {Universality in the
  tripartite information after global quenches: (generalised) quantum xy
  models},\ }\href {https://doi.org/10.1007/JHEP06(2023)140} {\bibfield
  {journal} {\bibinfo  {journal} {Journal of High Energy Physics}\ }\textbf
  {\bibinfo {volume} {2023}},\ \bibinfo {pages} {140} (\bibinfo {year}
  {2023})}\BibitemShut {NoStop}%
\bibitem [{\citenamefont {Alba}\ \emph {et~al.}(2021)\citenamefont {Alba},
  \citenamefont {Bertini}, \citenamefont {Fagotti}, \citenamefont {Piroli},\
  and\ \citenamefont {Ruggiero}}]{Alba2021Generalized}%
  \BibitemOpen
  \bibfield  {author} {\bibinfo {author} {\bibfnamefont {V.}~\bibnamefont
  {Alba}}, \bibinfo {author} {\bibfnamefont {B.}~\bibnamefont {Bertini}},
  \bibinfo {author} {\bibfnamefont {M.}~\bibnamefont {Fagotti}}, \bibinfo
  {author} {\bibfnamefont {L.}~\bibnamefont {Piroli}},\ and\ \bibinfo {author}
  {\bibfnamefont {P.}~\bibnamefont {Ruggiero}},\ }\bibfield  {title} {\bibinfo
  {title} {{Generalized-hydrodynamic approach to inhomogeneous quenches:
  correlations, entanglement and quantum effects}},\ }\href
  {https://doi.org/10.1088/1742-5468/ac257d} {\bibfield  {journal} {\bibinfo
  {journal} {Journal of Statistical Mechanics: Theory and Experiment}\ }\textbf
  {\bibinfo {volume} {2021}},\ \bibinfo {pages} {114004} (\bibinfo {year}
  {2021})}\BibitemShut {NoStop}%
\bibitem [{\citenamefont {Mari\'c}()}]{Maric2023universality2}%
  \BibitemOpen
  \bibfield  {author} {\bibinfo {author} {\bibfnamefont {V.}~\bibnamefont
  {Mari\'c}},\ }\bibfield  {title} {\bibinfo {title} {Universality in the
  tripartite information after global quenches: spin flip and semilocal
  charges},\ }\Eprint {https://arxiv.org/abs/2307.01842} {arXiv:2307.01842
  [cond-mat.stat-mech]} \BibitemShut {NoStop}%
\bibitem [{\citenamefont {Alba}\ \emph {et~al.}(2009)\citenamefont {Alba},
  \citenamefont {Fagotti},\ and\ \citenamefont
  {Calabrese}}]{Alba2009Entanglement}%
  \BibitemOpen
  \bibfield  {author} {\bibinfo {author} {\bibfnamefont {V.}~\bibnamefont
  {Alba}}, \bibinfo {author} {\bibfnamefont {M.}~\bibnamefont {Fagotti}},\ and\
  \bibinfo {author} {\bibfnamefont {P.}~\bibnamefont {Calabrese}},\ }\bibfield
  {title} {\bibinfo {title} {{Entanglement entropy of excited states}},\ }\href
  {https://doi.org/10.1088/1742-5468/2009/10/p10020} {\bibfield  {journal}
  {\bibinfo  {journal} {Journal of Statistical Mechanics: Theory and
  Experiment}\ }\textbf {\bibinfo {volume} {2009}},\ \bibinfo {pages} {P10020}
  (\bibinfo {year} {2009})}\BibitemShut {NoStop}%
\bibitem [{\citenamefont {Ares}\ \emph {et~al.}(2014)\citenamefont {Ares},
  \citenamefont {Esteve}, \citenamefont {Falceto},\ and\ \citenamefont
  {S{\'{a}}nchez-Burillo}}]{Ares2014}%
  \BibitemOpen
  \bibfield  {author} {\bibinfo {author} {\bibfnamefont {F.}~\bibnamefont
  {Ares}}, \bibinfo {author} {\bibfnamefont {J.~G.}\ \bibnamefont {Esteve}},
  \bibinfo {author} {\bibfnamefont {F.}~\bibnamefont {Falceto}},\ and\ \bibinfo
  {author} {\bibfnamefont {E.}~\bibnamefont {S{\'{a}}nchez-Burillo}},\
  }\bibfield  {title} {\bibinfo {title} {Excited state entanglement in
  homogeneous fermionic chains},\ }\href
  {https://doi.org/10.1088/1751-8113/47/24/245301} {\bibfield  {journal}
  {\bibinfo  {journal} {Journal of Physics A: Mathematical and Theoretical}\
  }\textbf {\bibinfo {volume} {47}},\ \bibinfo {pages} {245301} (\bibinfo
  {year} {2014})}\BibitemShut {NoStop}%
\bibitem [{\citenamefont {Yang}\ and\ \citenamefont
  {Yang}(1969)}]{Yang1969Thermodynamics}%
  \BibitemOpen
  \bibfield  {author} {\bibinfo {author} {\bibfnamefont {C.~N.}\ \bibnamefont
  {Yang}}\ and\ \bibinfo {author} {\bibfnamefont {C.~P.}\ \bibnamefont
  {Yang}},\ }\bibfield  {title} {\bibinfo {title} {Thermodynamics of a
  one‐dimensional system of bosons with repulsive delta‐function
  interaction},\ }\href {https://doi.org/10.1063/1.1664947} {\bibfield
  {journal} {\bibinfo  {journal} {Journal of Mathematical Physics}\ }\textbf
  {\bibinfo {volume} {10}},\ \bibinfo {pages} {1115} (\bibinfo {year}
  {1969})},\ \Eprint {https://arxiv.org/abs/https://doi.org/10.1063/1.1664947}
  {https://doi.org/10.1063/1.1664947} \BibitemShut {NoStop}%
\bibitem [{\citenamefont {Korepin}\ \emph {et~al.}(1993)\citenamefont
  {Korepin}, \citenamefont {Bogoliubov},\ and\ \citenamefont
  {Izergin}}]{korepin_bogoliubov_izergin_1993}%
  \BibitemOpen
  \bibfield  {author} {\bibinfo {author} {\bibfnamefont {V.~E.}\ \bibnamefont
  {Korepin}}, \bibinfo {author} {\bibfnamefont {N.~M.}\ \bibnamefont
  {Bogoliubov}},\ and\ \bibinfo {author} {\bibfnamefont {A.~G.}\ \bibnamefont
  {Izergin}},\ }\href {https://doi.org/10.1017/CBO9780511628832} {\emph
  {\bibinfo {title} {Quantum Inverse Scattering Method and Correlation
  Functions}}},\ Cambridge Monographs on Mathematical Physics\ (\bibinfo
  {publisher} {Cambridge University Press},\ \bibinfo {year}
  {1993})\BibitemShut {NoStop}%
\bibitem [{\citenamefont {Mossel}\ and\ \citenamefont
  {Caux}(2012)}]{Mossel2012Generalized}%
  \BibitemOpen
  \bibfield  {author} {\bibinfo {author} {\bibfnamefont {J.}~\bibnamefont
  {Mossel}}\ and\ \bibinfo {author} {\bibfnamefont {J.-S.}\ \bibnamefont
  {Caux}},\ }\bibfield  {title} {\bibinfo {title} {{Generalized {TBA} and
  generalized Gibbs}},\ }\href {https://doi.org/10.1088/1751-8113/45/25/255001}
  {\bibfield  {journal} {\bibinfo  {journal} {Journal of Physics A:
  Mathematical and Theoretical}\ }\textbf {\bibinfo {volume} {45}},\ \bibinfo
  {pages} {255001} (\bibinfo {year} {2012})}\BibitemShut {NoStop}%
\bibitem [{\citenamefont {Cardy}\ \emph {et~al.}(2008)\citenamefont {Cardy},
  \citenamefont {Castro-Alvaredo},\ and\ \citenamefont
  {Doyon}}]{Cardy2008Form}%
  \BibitemOpen
  \bibfield  {author} {\bibinfo {author} {\bibfnamefont {J.~L.}\ \bibnamefont
  {Cardy}}, \bibinfo {author} {\bibfnamefont {O.~A.}\ \bibnamefont
  {Castro-Alvaredo}},\ and\ \bibinfo {author} {\bibfnamefont {B.}~\bibnamefont
  {Doyon}},\ }\bibfield  {title} {\bibinfo {title} {Form factors of
  branch-point twist fields in quantum integrable models and entanglement
  entropy},\ }\href {https://doi.org/10.1007/s10955-007-9422-x} {\bibfield
  {journal} {\bibinfo  {journal} {Journal of Statistical Physics}\ }\textbf
  {\bibinfo {volume} {130}},\ \bibinfo {pages} {129} (\bibinfo {year}
  {2008})}\BibitemShut {NoStop}%
\bibitem [{\citenamefont {Casini}\ and\ \citenamefont
  {Huerta}(2009{\natexlab{b}})}]{Casini2009Entanglement}%
  \BibitemOpen
  \bibfield  {author} {\bibinfo {author} {\bibfnamefont {H.}~\bibnamefont
  {Casini}}\ and\ \bibinfo {author} {\bibfnamefont {M.}~\bibnamefont
  {Huerta}},\ }\bibfield  {title} {\bibinfo {title} {{Entanglement entropy in
  free quantum field theory}},\ }\href
  {https://doi.org/10.1088/1751-8113/42/50/504007} {\bibfield  {journal}
  {\bibinfo  {journal} {Journal of Physics A: Mathematical and Theoretical}\
  }\textbf {\bibinfo {volume} {42}},\ \bibinfo {pages} {504007} (\bibinfo
  {year} {2009}{\natexlab{b}})}\BibitemShut {NoStop}%
\bibitem [{\citenamefont {Casini}\ and\ \citenamefont
  {Huerta}(2009{\natexlab{c}})}]{Casini2009reduced_density}%
  \BibitemOpen
  \bibfield  {author} {\bibinfo {author} {\bibfnamefont {H.}~\bibnamefont
  {Casini}}\ and\ \bibinfo {author} {\bibfnamefont {M.}~\bibnamefont
  {Huerta}},\ }\bibfield  {title} {\bibinfo {title} {Reduced density matrix and
  internal dynamics for multicomponent regions},\ }\href
  {https://doi.org/10.1088/0264-9381/26/18/185005} {\bibfield  {journal}
  {\bibinfo  {journal} {Classical and Quantum Gravity}\ }\textbf {\bibinfo
  {volume} {26}},\ \bibinfo {pages} {185005} (\bibinfo {year}
  {2009}{\natexlab{c}})}\BibitemShut {NoStop}%
\bibitem [{\citenamefont {Fries}\ and\ \citenamefont
  {Reyes}(2019)}]{Fries2019}%
  \BibitemOpen
  \bibfield  {author} {\bibinfo {author} {\bibfnamefont {P.}~\bibnamefont
  {Fries}}\ and\ \bibinfo {author} {\bibfnamefont {I.~A.}\ \bibnamefont
  {Reyes}},\ }\bibfield  {title} {\bibinfo {title} {Entanglement and relative
  entropy of a chiral fermion on the torus},\ }\href
  {https://doi.org/10.1103/PhysRevD.100.105015} {\bibfield  {journal} {\bibinfo
   {journal} {Phys. Rev. D}\ }\textbf {\bibinfo {volume} {100}},\ \bibinfo
  {pages} {105015} (\bibinfo {year} {2019})}\BibitemShut {NoStop}%
\bibitem [{\citenamefont {Blanco}\ \emph {et~al.}(2022)\citenamefont {Blanco},
  \citenamefont {Ferreira~Chase}, \citenamefont {Laurnagaray},\ and\
  \citenamefont {P\'erez-Nadal}}]{Blanco2022}%
  \BibitemOpen
  \bibfield  {author} {\bibinfo {author} {\bibfnamefont {D.}~\bibnamefont
  {Blanco}}, \bibinfo {author} {\bibfnamefont {T.}~\bibnamefont
  {Ferreira~Chase}}, \bibinfo {author} {\bibfnamefont {J.}~\bibnamefont
  {Laurnagaray}},\ and\ \bibinfo {author} {\bibfnamefont {G.}~\bibnamefont
  {P\'erez-Nadal}},\ }\bibfield  {title} {\bibinfo {title} {R\'enyi entropies
  of the massless dirac field on the torus},\ }\href
  {https://doi.org/10.1103/PhysRevD.105.045014} {\bibfield  {journal} {\bibinfo
   {journal} {Phys. Rev. D}\ }\textbf {\bibinfo {volume} {105}},\ \bibinfo
  {pages} {045014} (\bibinfo {year} {2022})}\BibitemShut {NoStop}%
\bibitem [{\citenamefont {Jaynes}(1957)}]{Jaynes1957}%
  \BibitemOpen
  \bibfield  {author} {\bibinfo {author} {\bibfnamefont {E.~T.}\ \bibnamefont
  {Jaynes}},\ }\bibfield  {title} {\bibinfo {title} {{Information Theory and
  Statistical Mechanics}},\ }\href {https://doi.org/10.1103/PhysRev.106.620}
  {\bibfield  {journal} {\bibinfo  {journal} {Phys. Rev.}\ }\textbf {\bibinfo
  {volume} {106}},\ \bibinfo {pages} {620} (\bibinfo {year}
  {1957})}\BibitemShut {NoStop}%
\bibitem [{\citenamefont {Rigol}\ \emph {et~al.}(2007)\citenamefont {Rigol},
  \citenamefont {Dunjko}, \citenamefont {Yurovsky},\ and\ \citenamefont
  {Olshanii}}]{Rigol2007Relaxation}%
  \BibitemOpen
  \bibfield  {author} {\bibinfo {author} {\bibfnamefont {M.}~\bibnamefont
  {Rigol}}, \bibinfo {author} {\bibfnamefont {V.}~\bibnamefont {Dunjko}},
  \bibinfo {author} {\bibfnamefont {V.}~\bibnamefont {Yurovsky}},\ and\
  \bibinfo {author} {\bibfnamefont {M.}~\bibnamefont {Olshanii}},\ }\bibfield
  {title} {\bibinfo {title} {{Relaxation in a Completely Integrable Many-Body
  Quantum System: An Ab Initio Study of the Dynamics of the Highly Excited
  States of 1D Lattice Hard-Core Bosons}},\ }\href
  {https://doi.org/10.1103/PhysRevLett.98.050405} {\bibfield  {journal}
  {\bibinfo  {journal} {Phys. Rev. Lett.}\ }\textbf {\bibinfo {volume} {98}},\
  \bibinfo {pages} {050405} (\bibinfo {year} {2007})}\BibitemShut {NoStop}%
\bibitem [{\citenamefont {Vidal}\ \emph {et~al.}(2003)\citenamefont {Vidal},
  \citenamefont {Latorre}, \citenamefont {Rico},\ and\ \citenamefont
  {Kitaev}}]{Vidal2003Entanglement}%
  \BibitemOpen
  \bibfield  {author} {\bibinfo {author} {\bibfnamefont {G.}~\bibnamefont
  {Vidal}}, \bibinfo {author} {\bibfnamefont {J.~I.}\ \bibnamefont {Latorre}},
  \bibinfo {author} {\bibfnamefont {E.}~\bibnamefont {Rico}},\ and\ \bibinfo
  {author} {\bibfnamefont {A.}~\bibnamefont {Kitaev}},\ }\bibfield  {title}
  {\bibinfo {title} {{Entanglement in Quantum Critical Phenomena}},\ }\href
  {https://doi.org/10.1103/PhysRevLett.90.227902} {\bibfield  {journal}
  {\bibinfo  {journal} {Phys. Rev. Lett.}\ }\textbf {\bibinfo {volume} {90}},\
  \bibinfo {pages} {227902} (\bibinfo {year} {2003})}\BibitemShut {NoStop}%
\bibitem [{\citenamefont {Jin}\ and\ \citenamefont {Korepin}(2004)}]{Jin2004}%
  \BibitemOpen
  \bibfield  {author} {\bibinfo {author} {\bibfnamefont {B.-Q.}\ \bibnamefont
  {Jin}}\ and\ \bibinfo {author} {\bibfnamefont {V.~E.}\ \bibnamefont
  {Korepin}},\ }\bibfield  {title} {\bibinfo {title} {{Quantum Spin Chain,
  Toeplitz Determinants and the Fisher---Hartwig Conjecture}},\ }\href
  {https://doi.org/10.1023/B:JOSS.0000037230.37166.42} {\bibfield  {journal}
  {\bibinfo  {journal} {Journal of Statistical Physics}\ }\textbf {\bibinfo
  {volume} {116}},\ \bibinfo {pages} {79} (\bibinfo {year} {2004})}\BibitemShut
  {NoStop}%
\end{thebibliography}%
\end{document}